\documentclass[prl,twocolumn,showpacs,amsmath,amssymb]{revtex4}

\usepackage{graphicx}
\usepackage{dcolumn}
\usepackage{bm}

\begin{document}


\title{Symmetric and non-symmetric vortex-antivortex molecules\\ in fourfold superconducting geometry}

\author{R. Geurts}
\author{M. V. Milo\v{s}evi\'{c}}
\author{F. M. Peeters}
\email{francois.peeters@ua.ac.be}

\affiliation{Departement Fysica, Universiteit Antwerpen,
Groenenborgerlaan 171, B-2020 Antwerpen, Belgium}

\date{\today}

\begin{abstract}
In submicron superconducting squares in a homogeneous magnetic
field, Ginzburg-Landau theory may admit solutions of the
vortex-antivortex type, conforming with the symmetry of the sample
[Chibotaru {\it et al.}, Nature {\bf 408}, 833 (2000)]. Here we
show that these fascinating, but never experimentally observed
states, can be {\it enforced by artificial fourfold pinning}, with
their diagnostic features {\it enhanced by orders of magnitude}.
The second-order nucleation of vortex-antivortex molecules can be
driven either by temperature or applied magnetic field, with
stable {\it asymmetric vortex-antivortex equilibria} found on its
path.
\end{abstract}


\maketitle

All sub-atomic particles have a fraternal twin: an antiparticle,
exactly alike except for e.g. opposite charge or helicity.
However, as a general rule, matter and antimatter cannot coexist
without excess energy and annihilate each other.

This universal duality has its analogue in the physics of
superconductors, where vortices as the carriers of magnetic flux
play the role of charged particles. Namely, vortex-antivortex
pairs in superconducting films can be induced in a local hot-spot,
created by thermal fluctuations \cite{BKT}, photon absorption
\cite{PHO}, or driving current \cite{DCR}. Intuitively,
vortex-antivortex pairs are easily stabilized in an inhomogeneous
magnetic field, such as one resulting from dipolar magnetic
objects in the vicinity of the superconductor \cite{LeuvMisko}.

Contrary to latter examples, recent theoretical studies
\cite{VAV,VAV1,VAV2} have shown that vortex-antivortex structures
can be stabilized in submicron superconductors even in {\it
homogeneous} magnetic field, and {\it without any apparent energy
input}. These findings are restricted to the cases when vortex
structure does not conform with the sample geometry. A typical
example is the C$_3$ symmetry of the three-vortex state in a
superconducting square; in spite of the unipolar applied field,
the C$_4$ state of four vortices with a central antivortex may
become energetically preferable. Similarly to the solid-fluid and
the ferromagnetic symmetry-breaking processes, the more
symmetrical phase is found on the high-temperature side of the
transition, and the less symmetrical one on the low-temperature
side. Landau pointed out that one can always unequivocally
determine whether or not given state possesses a given symmetry
\cite{and}. Therefore, it is not possible to analytically deform a
state in one phase into a phase possessing truly different
symmetry. This means, for example, that it is impossible for the
solid-liquid phase boundary to end in a critical point like the
liquid-gas boundary. Nevertheless, the above described
vortex-antivortex nucleation during the C$_3$ to C$_4$ transition
in mesoscopic superconductors is of second-order with respect to
temperature. The appearance and stability of these {\it asymmetric
vortex-antivortex molecules} in the ground state is one of the
main objectives of this Letter.

It should be noted here that the symmetry-induced
vortex-antivortex configurations remain experimentally undetected,
mainly because of their high sensitivity to defects in sample
edges \cite{VAV1}. As another drawback, due to extreme vortex
proximity \cite{VAV2}, those states are undistinguishable from a
single multiquanta vortex for conventional techniques such as
scanning-tunnelling and Hall-probe microscopy. Theory suggests
that the latter problem can be solved by a magnetic dot deposited
on the sample \cite{VAV3}. Added inhomogeneous magnetic potential
interacts dually with present vortex and antivortex
\cite{MiskoOut} and separates them in a controllable fashion, but
disturbs the conceptual novelty of no external stabilizing factor
added to the vortex-antivortex coexistence.
\begin{figure}[b]
\includegraphics[width=5.cm]{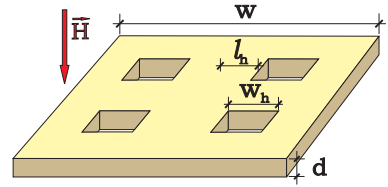}
\caption{\label{fig1} Superconducting square (size $w\times w$ and
thickness $d$) with four nanoholes (each $w_h\times w_h$ in size,
displaced by $l_h$ from $x$ and $y$ symmetry lines). Indicated
direction of applied homogeneous magnetic field $H$ is denoted as
positive.}
\end{figure}

To address above issues, we discuss in this Letter the properties
and the improved observation conditions of a vortex-antivortex
state in a square superconducting sample {\it with fourfold
pinning}. We introduce a cluster of $2\times 2$ nanoholes to
enforce the C$_4$ symmetry, the non-energetic source of
vortex-antivortex nucleation (see Fig. \ref{fig1}). Albeit,
induced states {\it do not always obey the imposed symmetry}. We
show that similar symmetry breaking can be achieved in a classical
cluster of oppositely charged particles in a strongly confined
geometry. The latter setup has recently emerged as the `standard
model' for a variety of systems on different energy and length
scales, not only flux lines in superconductors, but also vortices
in Bose-Einstein condensates, electrons on liquid helium,
colloidal suspensions, and dust particles in plasmas.

Our theoretical approach relies upon the Ginzburg-Landau (GL)
theory, where we solve self-consistently a set of mean field
differential equations for the order parameter $\psi$ and the
vector potential ${\bf A}$
\begin{eqnarray}
&& (-i\nabla-{\bf A})^2\psi = (1-T-|\psi|^2)\psi, \label{gl1}\\
&& -\kappa^2\nabla\times\nabla\times {\bf A} = {\bf j}~.
\label{gl2}
\end{eqnarray}
The latter is the Maxwell-Amp\`{e}re equation with a current
density ${\bf j}=\Im(\psi^*\nabla\psi)-|\psi|^2{\bf A}$. $\kappa$
is the material property and equals the ratio between the magnetic
field penetration depth $\lambda$ and coherence length $\xi$.
Solution of Eqs. (\ref{gl1}, \ref{gl2}) minimizes Gibbs free
energy
$\mathcal{G}/\mathcal{G}_0=\Omega^{-1}\int{[-|\psi|^4+2({\bf
A}-{\bf A}_0)\cdot {\bf j}]d\Omega}$, where the condition
$j_\perp=0$ was used on the boundaries of the superconducting
volume $\Omega$, and $\mathcal{G}_0$ stands for the
superconducting condensation energy $H_{c}^{2}\big/4\pi$. In above
expressions, $\nabla\times{\bf A}_0={\bf H}$ denotes the applied
magnetic field, and $\nabla\times{\bf A}={\bf h}$ the resulting
local field in the sample. All distances are expressed in units of
$\xi_0=\xi$(T=0), the vector potential in $\phi_0\big/2\pi\xi_0$,
and the order parameter in $\sqrt{-\alpha /\beta}$ with $\alpha $,
$\beta $ being the GL coefficients. For details of the numerics we
refer to Ref. \cite{schw}.

\begin{figure}[t]
\includegraphics[width=240pt]{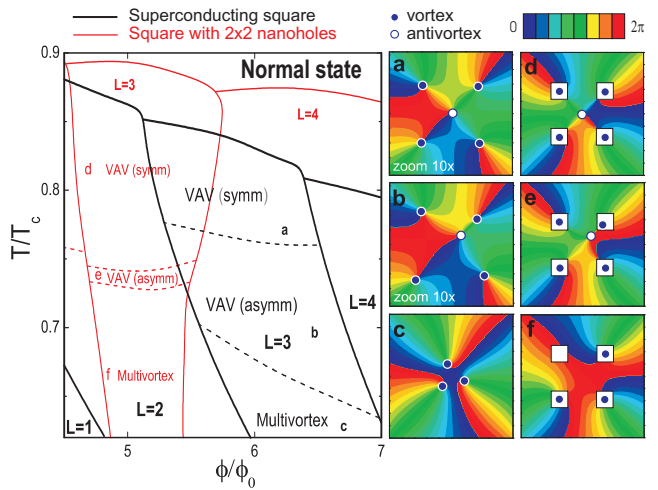}
\caption{\label{fig2} Temperature versus magnetic field
equilibrium phase diagram. Solid lines illustrate $1^{st}$ order,
and dashed $2^{nd}$ order transitions in the ground state.
Parameters used: $\kappa = 1$, $d = 1 \xi_0$, $w=10\xi_0$, $l_h/w
= w_h/w = 0.125$ (see Fig. \ref{fig1}). Contourplots (a)-(f) show
the phase of the order parameter of particular vortex states (as
indicated in the left diagram).}
\end{figure}

\paragraph{\bf Symmetry-induced asymmetry}
Fig. \ref{fig2} shows the ground-state phase diagram of our
perforated sample, compared to the previously studied one without
holes \cite{VAV,VAV1,VAV2,VAV3,slava}. Namely, at given
temperature $T$ (scaled to critical temperature $T_c$), we obtain
stable solutions of Eqs. (\ref{gl1}, \ref{gl2}) by starting the
iterative procedure from randomly generated initial conditions,
while applied magnetic field $H$ is swept up/down. The ground
state is then determined by comparing the energy of all found
states. Note that dimensions of the system are given in $\xi_0$,
so that depending on the material our results apply to samples
from hundred nanometers to several microns in size. Parameter
$\kappa$ is taken equal 1.

The size-field phase diagram [size scaled to
$\xi(T)=\xi_0/\sqrt{1-(T/T_c)^2}$] for a square sample without
holes has been studied previously in Refs. \cite{VAV,VAV2}. Since
in Ref. \cite{VAV} the linearized GL theory is used, valid only
extremely close to the superconducting/normal (S/N) phase
boundary, those results were quantitatively corrected in Ref.
\cite{VAV2} where both GL equations were numerically solved.
However, authors kept parameter $\kappa_{eff}=\kappa^2\xi(T)/d$
fixed when temperature was changed, so their diagram did not
reflect true temperature dependence of the vortex state.
Nevertheless, our phase diagram (with temperature dependence taken
explicitly) shows similar qualitative behavior - sets of
individual vortices are found at lower temperatures, which merge
into a giant vortex at higher temperatures for all vorticities (L)
except $L=3$. The triangular symmetry of that state [Fig.
\ref{fig2}(c)] is not favorable closer to $T_c$ when square
confinement dominates, and the L=4-1(=3) vortex-antivortex state
is induced [Fig. \ref{fig2}(a)].

However, contrary to previous works, our analysis shows that the
vortex-antivortex (VAV) nucleation process is of second order
(dashed lines in Fig. \ref{fig2}) over a {\it broad temperature
and field range}. In other words, with changing applied magnetic
field ($\Delta\phi\approx \phi_0$) or temperature ($\Delta
T\approx 0.1T_c$) while in the $L=3$ state, besides the three
existing vortices {\it a vortex-antivortex pair gradually
dissociates}. A remarkable non-symmetric $L=4-1$ state is induced
[Fig. \ref{fig2}(b)], and {\it the ground-state exhibits symmetry
breaking} during the transition between intrinsically symmetric
states (C$_3$ vs. C$_4$).

Although these asymmetric equilibria seem fairly counterintuitive,
their stability follows from the competing interactions in this
system. In analogy, we recall the properties of small confined
clusters of particles, where the confinement would result from the
screening currents in our system (along the sample edges), and
charged particles represent vortices.

\paragraph{\bf Analogy to classical systems}
The model system was defined in Ref. \cite{bed}, where Hamiltonian
is given by $\mathcal{H}=\sum_{i>j}^{N}U_{ij}+\sum_{i}^{N}V(x,y)$,
with $N$ being the number of particles. We took square-parabolic
confinement potential of size $w_c$, i.e. $V(x,y) =
\frac{1}{2}m\omega^2_0\frac{x^2+y^2}{w^2_{c}}\left[1+\sqrt{\delta
+ \cos^2\left(2\arctan\frac{y}{x}\right)}\right]$ ($m$-mass of
particle, $\omega_0$-confinement frequency) with a nonzero $\delta
\ll 1$, ensuring the existence of the derivatives of this
potential in the corners of the square. To further translate this
system to vortices in mesoscopic superconductors, one has to
choose properly the vortex-(anti)vortex interaction energy
($U_{ij}$). Firstly, this interaction must be of long range (and
diminishing far from the source), knowing that supercurrents decay
$\sim 1/r$ away from the vortex. Secondly, $U_{ij}$ may not
diverge for $r_{ij}\leq \alpha$ but saturate, allowing for
realistic merging of vortices into a giant vortex \cite{kanda}, or
a vortex-antivortex annihilation. In this scenario, $\alpha$
roughly corresponds to the finite size of the vortex core.

Though any interaction energy of described profile would suit our
analysis, we used a modified logarithmic and modified Coulomb
interaction, more specifically
$dU/dr(r_{ij})=\beta(1-e^{-r_{ij}/\alpha})^2\big/r_{ij}$ and
$dU/dr(r_{ij})=\frac{q_iq_j}{\varepsilon}(1 -
e^{-r_{ij}/\alpha})^3\big/r_{ij}^2$, respectively.
Notice that the modified Coulomb interaction is a quality fit for
$\alpha^{-1}=\sqrt{2}\kappa$ (up to a multiplying constant) to the
realistic vortex-vortex interaction in type-II samples (see e.g.
Ref. \cite{brandt}).

\begin{figure}[t]
\includegraphics[width=7.5cm]{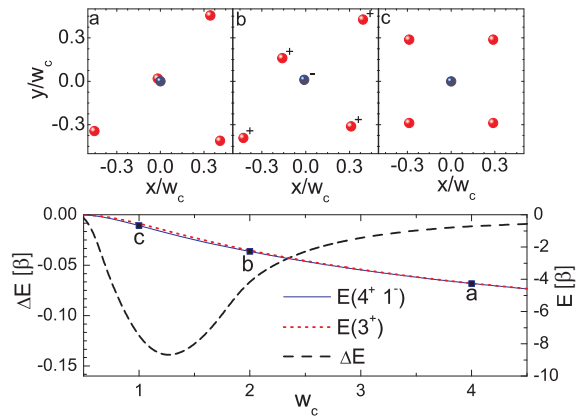}
\caption{\label{fig3} Energy of the 3-particle cluster compared to
the cluster of 4 particles and one antiparticle, as a function of
the size of the square confinement $w_c$, for modified logarithmic
inter-particle interaction (of strength $\beta$, and with
$\alpha=1$). (a-c) are snapshots of the ground state configuration
of the latter cluster with decreasing $w_c$ ($x,y$ in units of
$\sqrt{\frac{2\beta}{m\omega^2_0}}$).}
\end{figure}

Next we employed molecular dynamics simulation, a relatively
simple and rapidly convergent technique that gives a reliable
estimate of the energy for small clusters of interacting
particles. The obtained energy versus confinement size ($w_c$) is
shown in Fig. \ref{fig3}, for a cluster containing three particles
of the same charge (i.e. $U_{ij}>0$) and a cluster of four
particles and an antiparticle, with clear analogy to our $L=3$ and
$L=4-1$ vortex state. For large $w_c$, the ground states of the
two clusters have identical energy, as one particle and the
antiparticle sit on top of each other. However, for tighter
confinement, the $N=4^+ 1^-$ cluster {\it attains lower energy}
than the $N=3^+$ one. Namely, particle and antiparticle gradually
separate causing the rearrangement of the remaining 3 particles.
Figs. \ref{fig3}(a-c) show snapshots of this dynamical transition
in the ground state. To our knowledge, this symmetry breaking in
the crystallization of confined systems is novel and applies to
classical particles interacting with charged impurities in the
substrate \cite{farias}, and potentially to small ionic crystals
under pressure \cite{joch}. Note also that colloidal molecules
have recently been realized experimentally on hydrophilic square
templates \cite{coll}.
\begin{figure}[b]
\includegraphics[width=7.5cm]{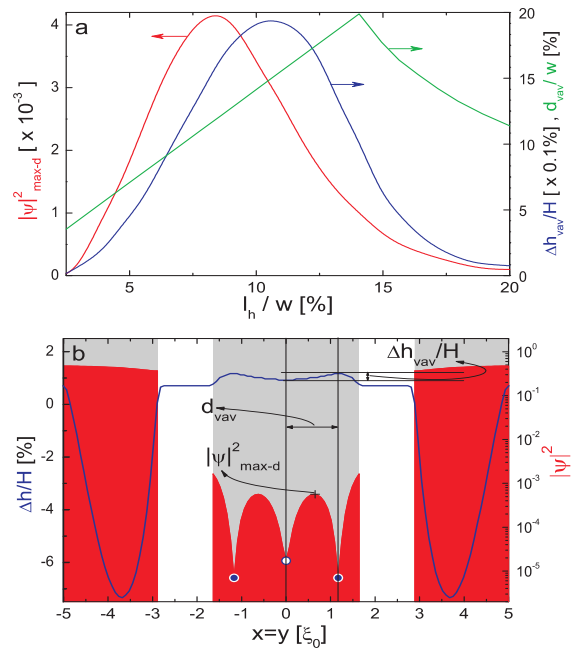}
\caption{\label{fig4} The properties of the $L=4-1$
vortex-antivortex state in the perforated sample, at $T=0.84T_c$,
and applied flux $\phi=5.5\phi_0$: (a) the distance, maximal
$|\Psi|^2$-density, and difference in the magnetic response
between vortex and antivortex, as a function of the position of
the holes; (b) the Cooper-pair density and the magnetic field
profile ($\Delta h=h-H$) along the diagonal of the sample, in case
when vortices are not residing in holes [$l_h=0.165w$, see (a)].}
\end{figure}

\paragraph{\bf Distinction of antivortex in fourfold pinning}
As shown in Fig. \ref{fig2}, in our sample with nanoholes, the
symmetrically placed perforations (i) pin (attract) vortices
individually, i.e. favor the multivortex state, and (ii) stabilize
vortex states commensurate with the square geometry, e.g. $L=2$
and particularly $L=4$ \cite{golib}. Although $L=3$ state is
somewhat suppressed in the $H-T$ space compared to the plain
square case, the antivortex in the L=4-1 state {\it is clearly
pronounced} [see Fig. \ref{fig2}(d,e)] since remaining 4 vortices
are captured by the holes. To illustrate better this issue, we
calculated the vortex-antivortex distance ($d_{vav}$) and
amplitudes of the Cooper-pair density and local magnetic field
between them, as a function of the sample parameters. As shown in
Fig. \ref{fig4}, we are able to achieve the vortex-antivortex
distance of 5-20\% of the sample size, which {\it greatly exceeds}
the distances found in plain squares (see Ref. \cite{VAV2}),
especially in larger samples. At the same time, the amplitudes of
both order parameter and local magnetic field are {\it enhanced by
almost two orders of magnitude} compared to previous studies, and
reach the limits of conventional scanning techniques. Our study
shows that $w_h=l_h=12.5\%w$ are optimal parameters for the
observation of the antivortex, as larger holes capture more
applied flux and decrease the field in the center of the sample,
whereas larger distance between the holes results in a weaker
influence on the vortex-antivortex state in the sample center (see
Fig. \ref{fig4}). It should also be emphasized that fourfold
distribution of holes is more important than the actual shape of
the sample, i.e. {\it even a circular disk with four symmetrically
arranged holes exhibits the $L=4-1$ state}.

One should note that the inter-vortex interaction is governed by
$\kappa$, and may even change sign in type-I samples
\cite{brandt}. This strongly affects the vortex configurations,
and consequently the vortex-antivortex stabilization. In our
calculation for both squares with and without nanoholes, decrease
of $\kappa$ {\it did not favor the $L=4-1$ state} but instead the
$L=3$ giant vortex becomes more stable, contrary to the findings
of Ref. \cite{slava}. Apparently, suggested vortex-antivortex
repulsion at lower $\kappa$ is dominated by attraction between the
remaining vortices. On the other hand, for larger $\kappa$, the
shorter range of inter-vortex interactions reduces the stability
of asymmetric states which rely on strong competing forces.
Therefore, material properties are very important for potential
experimental observation of the antivortex, as $\kappa$ can be
arguably controlled by impurities, which increase $\lambda$ and
decrease $\xi$.

\begin{figure}[t]
\includegraphics[width=240pt]{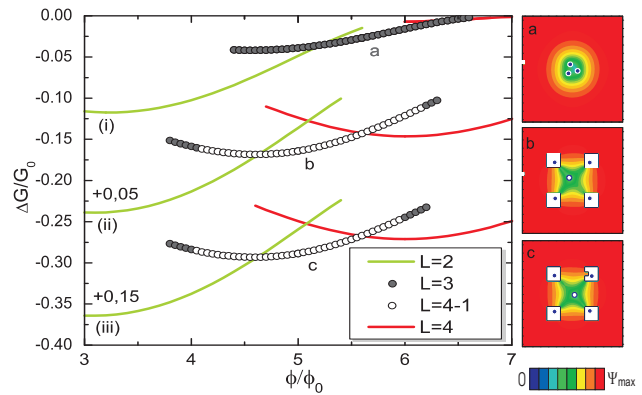}
\caption{\label{fig5} Gibbs free energy for $T=0.78T_c$ and (i)
plain square sample with an edge-defect (size $0.5\xi_0$), (ii)
sample with 2x2 holes and an edge-defect, and (iii) sample with
2x2 holes and a defect in hole-edges. Insets (a-c) show the
Cooper-pair density plots of found states with total vorticity 3,
in cases (i-iii).}
\end{figure}

\paragraph{\bf Enhanced stamina of vortex-antivortex}
Yet another crucial experimental issue is the sensitiveness of the
vortex-antivortex state to surface defects. While the ground-state
strongly depends on symmetry, the analysis of Ref. \cite{VAV1} has
shown that defects as small as $0.01w$ at the edges of a plain
superconducting square disable vortex-antivortex nucleation. We
found that $2\times 2$ nanoholes significantly strengthen the
$L=4-1$ state. Fig. \ref{fig5} shows the free energy diagrams of
(i) a plain square with a small edge defect, (ii) a square with
nanoholes and an edge defect, and (iii) square with nanoholes and
a defect at one of the holes. Whereas in case (i) {\it no}
vortex-antivortex state is found, in both cases (ii) and (iii) we
found stable $L=4-1$ state. Our numerical experiment shows that
vortex-antivortex states in latter cases can survive for defects
in edges {\it up to $10\%$ of the sample size} and defects in
nanoholes up to {\it remarkable $40\%$} of their size. Note also
that the presence of defects changes the current profile of the
sample, which again results in asymmetric vortex-antivortex
configurations [see Fig. \ref{fig5}(b,c)].

Therefore, we may conclude that artificial fourfold pinning
vigorously enhances (in all relevant aspects) the experimental
observability of vortex-antivortex molecules in flat
superconducting samples. The verification of the found asymmetric
vortex-antivortex equilibria is of particular importance, as this
symmetry-breaking is predicted both for many-body and few-body
systems with strong competing interactions in polygonal
constraints.

This work was supported by the Flemish Science Foundation
(FWO-Vl), the Belgian Science Policy, the JSPS/ESF-NES program,
and the ESF-AQDJJ network.

\end{document}